\documentstyle[epsf]{article}
\newcommand{\gsim}{\mbox{\raisebox{-.3em}{$\stackrel{>}{\sim}$}}}
\newcommand{\lsim}{\mbox{\raisebox{-.3em}{$\stackrel{<}{\sim}$}}}

\renewcommand{\cite}[1]{\ref{#1}}

\newcommand{\beq}{\begin{equation}}
\newcommand{\eeq}{\end{equation}}
\newcommand{\beqa}{\begin{eqnarray}}
\newcommand{\eeqa}{\end{eqnarray}}
\newcommand{\bpr}{\begin{problem}}
\newcommand{\epr}{\end{problem}}
\newcommand{\bcent}{\begin{center}}
\newcommand{\ecent}{\end{center}}
\newcommand{\bfig}{\begin{figure}}
\newcommand{\efig}{\end{figure}}
\newcommand{\bpc}{\begin{picture}}
\newcommand{\epc}{\end{picture}}

\newcommand{\nnb}{\nonumber}
\newcommand{\reflef}{(\ref}
\newcommand{\MP}{M_{\rm P}}

\begin{document} 
\baselineskip=0.6cm

\bcent
{\LARGE\bf Varying Fine-Structure Constant and  the Cosmological Constant Problem}\footnote{Delivered at JENAM 2002, Porto, Portugal, 2-7 September 2002, to be published in Proceedings.}\\[.4em]
Yasunori Fujii\footnote{E-mail: fujii@dd.catv.ne.jp}
\ecent
%\mbox{}\\[.0em]
\baselineskip=0.54cm
\noindent
\bcent
{\large\bf Abstract}\\[1.8em]
\begin{minipage}{13cm}
{\small
We start with a brief account of the latest analysis of the Oklo phenomenon providing the still most stringent constraint on time-variability of the fine-structure constant $\alpha$.  Comparing this with the recent result from the measurement of distant QSO's appears to indicate a non-uniform time-dependence, which we argue to be related to another recent finding of the accelerating universe. This view is implemented in terms of the scalar-tensor theory, applied specifically to the small but nonzero cosmological constant.  Our detailed calculation shows that these two phenomena can be understood in terms of a common origin, a particular behavior of the scalar field, dilaton.  We also sketch how this theoretical approach makes it appropriate to revisit non-Newtonian gravity featuring small violation of Weak Equivalence Principle at medium distances.
}
\end{minipage}
\ecent
%\mbox{}

%\keywords{Oklo phenomenon, cosmological constant, varying fine-structure constant}

%%%%%%%%%%%%%%%%%%%%%%%%%%%%%%%%%%%%%%%%%%%%%
\section{Constraint from Oklo}
%%%%%%%%%%%%%%%%%%%%%%%%%%%%%%%%%%%%%%%%%%%%%

Around 1974 it was finally agreed that self-sustained fission reactions took place {\em naturally} in Oklo, Gabon, some 2 billion years ago.  This is the ``Oklo phenomenon."  Shlyakhter [\cite{shly}] came to notice that measuring isotopic ratio of $^{149}{\rm Sm}$ left in the remnants of the ``natural reactors" is useful to determine how much nuclear phenomena 2 billion years ago could have been different from what they are. He focused on the reaction $n+^{149}\hspace{-.2em}{\rm Sm} \rightarrow ^{150}\hspace{-.2em}{\rm Sm} + \gamma$, which is unique in that  it is dominated by a resonance lying  as low as 97.3 meV, nearly 7 orders of magnitude too small compared with MeV, a typical energy scale of nuclear physics.  A very small number like this should be due to a nearly complete cancellation between two large effects; a repulsive Coulomb force proportional to $\alpha$ and the attractive nuclear force.

The resonance shows up as a sharp peak in the cross section plotted against a possible change $\Delta E_{\rm r}$ of the resonance energy $E_{\rm r}$.  We also assume thermal equilibrium of the neutron flux.  Even a slight change of $\alpha$ may result in the sizable change of $E_{\rm r}$ then of the cross section.  Thanks to this {\em amplification mechanism},  he derived the upper bound; $|\dot{\alpha}/\alpha|\lsim 10^{-17}{\rm y}^{-1}$, much better than any other results for years.

Unfortunately,  details of his derivation and the quality of the data have been left largely unknown.  However, our recent re-analysis [\cite{yf8}] shows that his ``champion result" itself can be trusted reasonably well.

To minimize the ``contamination" due to the inflow from {\em outside} the reactor that occurred {\em after} the end of reactor activity, we relied on the latest samples collected carefully from deep underground.  By analyzing the data on the isotopes of Sm, we computed the cross section $\hat{\sigma}_{149} = (91\pm 6) {\rm kb}$, as bounded by the two horizontal lines in Fig. 1.  Combining this with the shaded area corresponding to the improved estimate of the temperature $(200-400)^\circ{\rm C}$, we obtained the two intersections on both sides of the peak, then the two ranges:
%%%%%%%%%%%%%%%%%%%%%%%%%%
\beq
\Delta E_{\rm r} = \left\{ \begin{array}{rll}
 (9\pm 11 )&\hspace{-.6em}{\rm meV},\hspace{1em}&\mbox{Right-branch range,} \nnb\\[.5em]
 (-97\pm 8 )&\hspace{-.6em}{\rm meV}, \hspace{1em}&\mbox{Left-branch range.} \nnb
\end{array}
\right.
\label{Ebound}
\eeq
The right-branch range covers zero giving a ``null result" as usual, while the second one implies that the resonance energy 2 billion years ago was smaller than today's value beyond more than 12 $\sigma$.  

%%%%%%%%%%%%%%%%%%% fig. 1 %%%%%%%%%%%%%%%%%%%%%%%
\begin{figure}[t]
\hspace*{4.5cm}
\epsfxsize=5.cm
\epsffile{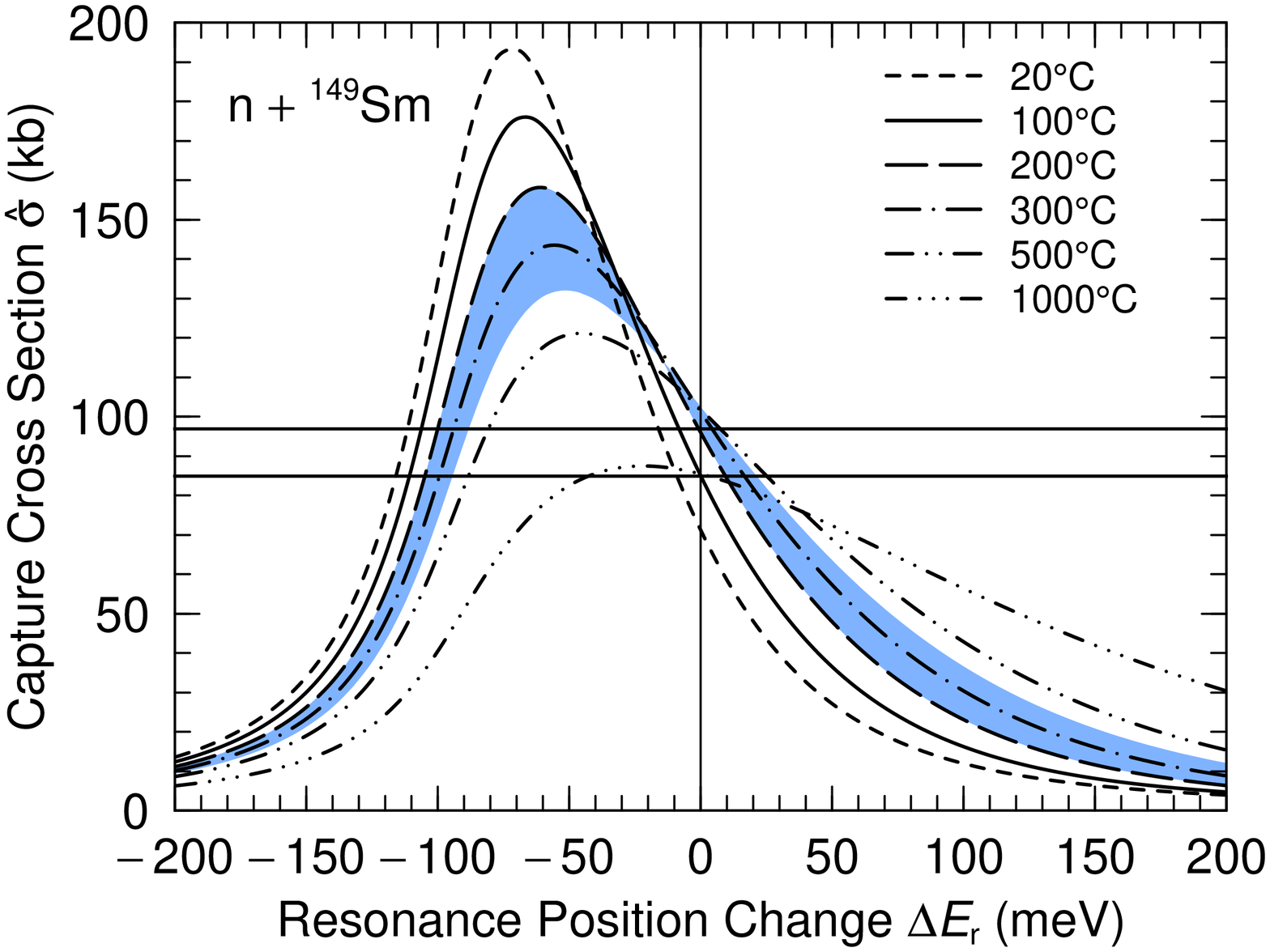}
\mbox{}\\
\hspace*{19.5em}{\scriptsize Fig. 1}

\end{figure}

We tried to see if the non-null result could be eliminated by looking at other isotopes ($^{155,157}{\rm Gd}, ^{113}{\rm Cd}$), but without a final conclusion yet.

We add that the same type of analysis due to Damour and Dyson [\cite{dd}] resulted in a single much more conservative bound as given by $-120{\rm meV} < \Delta E_r < 90{\rm meV}$.  They could have separated it into the narrower two, but then with the right-branch range failing to cover the zero, causing a suspicion of the contaminated effect of the data.

The bound \reflef{Ebound}) is translated into that of $\Delta\alpha$ by using $\Delta E_r = (\Delta\alpha /\alpha){\cal M}_c$ with ${\cal M}_c\approx -1.1{\rm MeV}$ estimated in [\cite{dd}], thus giving
%%%%%%%%%%%%%%%%%%%%%%%%%%%%%%%%%%
\beq
 \frac{\Delta \alpha}{\alpha} =
\left\{
\begin{array}{rll}
 -(0.8\pm 1.0)\times 10^{-8},\hspace{1em}&\mbox{Right-branch range,} \nnb\\[.6em]
 (0.88\pm 0.07) \times 10^{-7},\hspace{1em}&\mbox{Left-branch range}. \nnb
\end{array}
\right.
\label{Abound}
\eeq
The null-result, by dividing by $-2\times 10^9{\rm y}$,  gives the upper bound $\dot{\alpha}/\alpha = (0.4\pm 0.5)\times 10^{-17} {\rm y}^{-1}$, which ``happens" to agree reasonably well with Shlyakhter's result, as was mentioned.  But the agreement to this extent seems to be rather accidental, because, among other things, it is unlikely that the quality of the data available to him in the early years of the Oklo study was as good as ours.

We then move on to combine our result \reflef{Ebound}) with those non-null result obtained recently from QSO, by Webb {\it et al} [\cite{webb}]. The error bar of the Oklo data is almost invisible on the scale in the upper diagram of Fig. 2a.  It even appears as if $\alpha$ varies with time not necessarily in a simple manner.  We argue that the time-dependence might have something to do with the cosmological constant problem.

%%%%%%%%%%%%%%%%%% fig. 2 %%%%%%%%%%%%%%%%%%%%%%
\begin{figure}[t]
\hspace*{2.cm}
\parbox[b]{5.3cm}
{
\epsfxsize=5.0cm
\epsffile{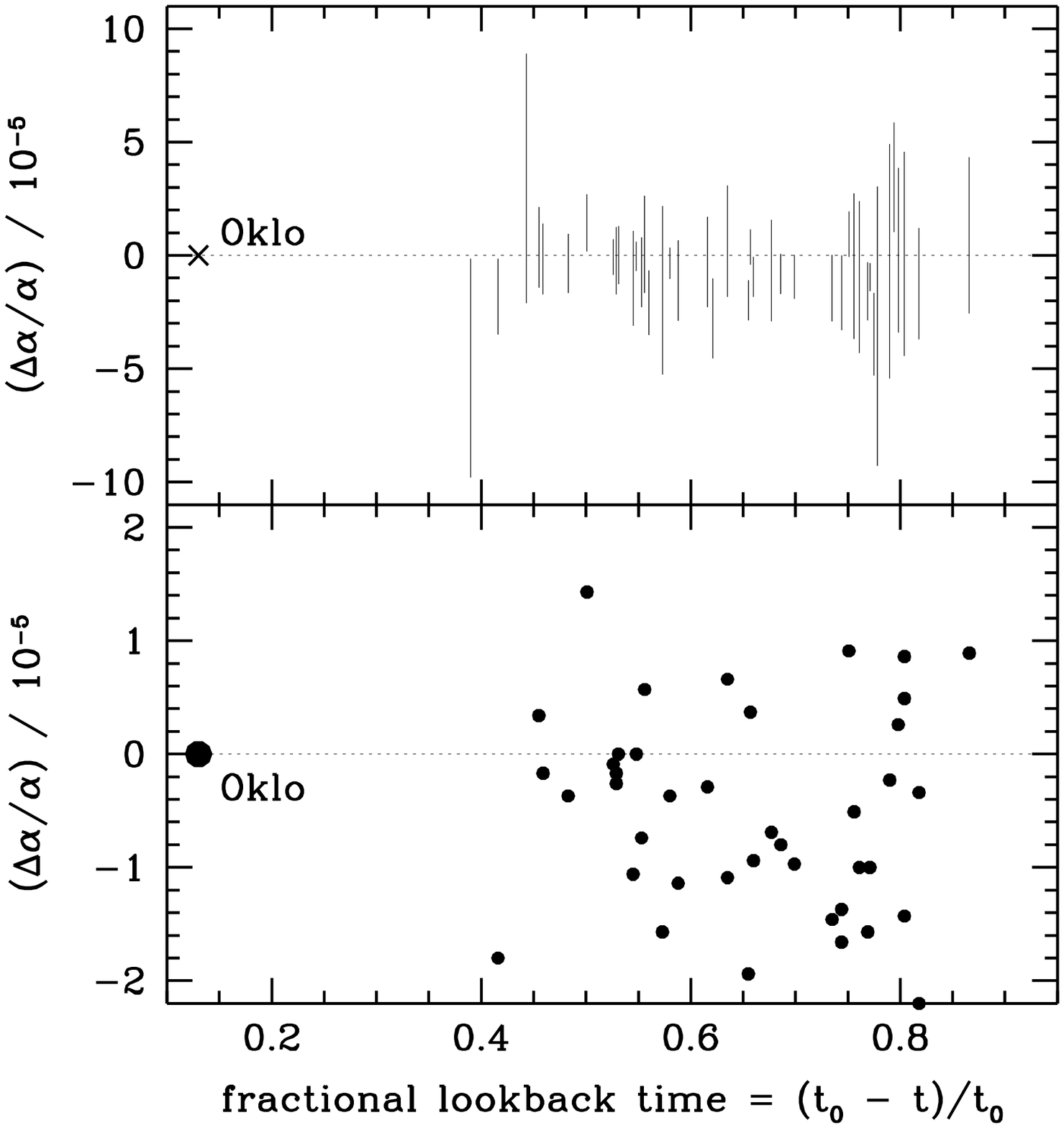}
\mbox{}\\[-1.7em]
\hspace*{6.0em}{\scriptsize Fig. 2a}
}
\parbox[b]{5.3cm}
{
\epsfxsize=5.0cm
\epsffile{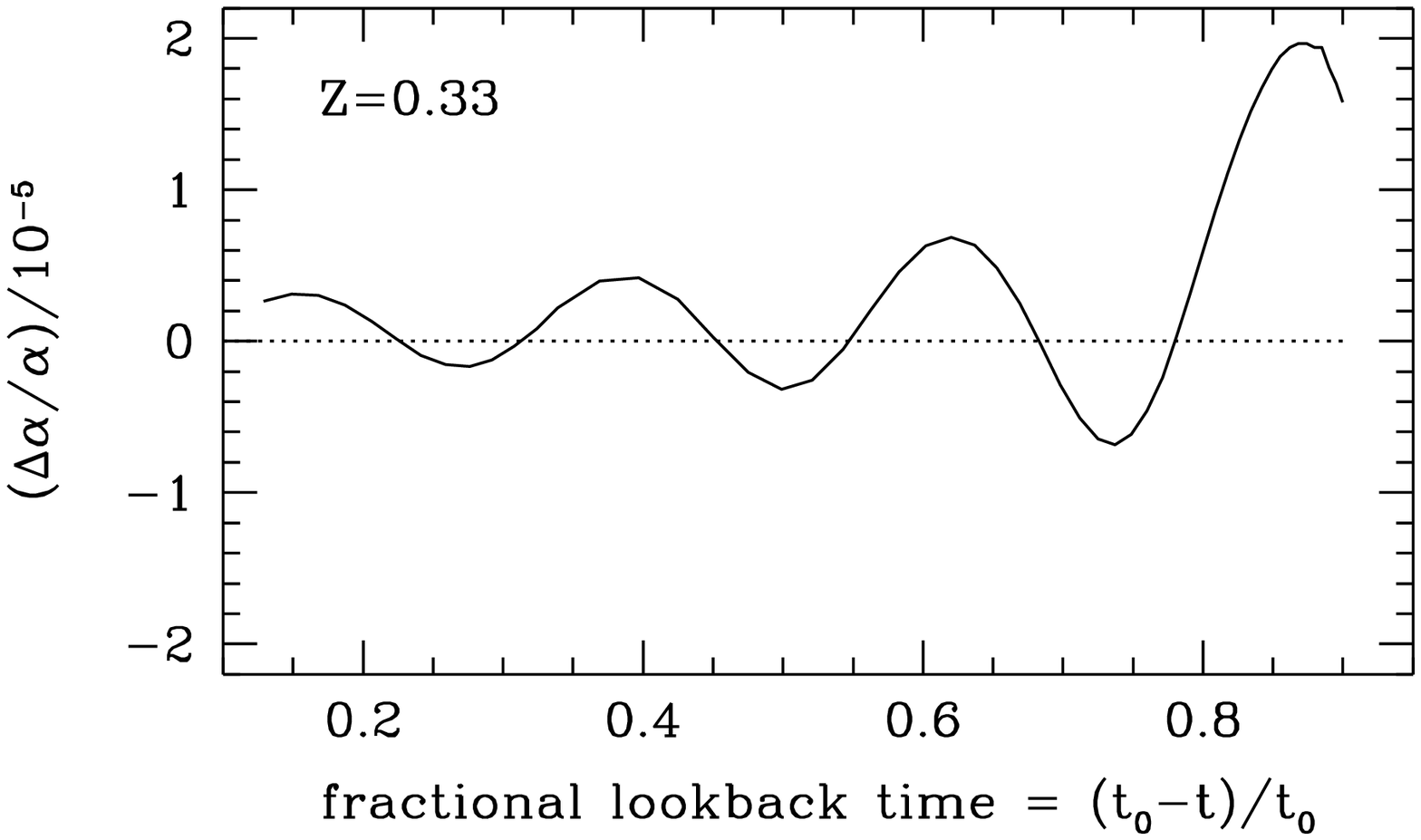}
\mbox{}\\[-1.6em]
\hspace*{6em}{\scriptsize Fig. 2b}
}
\end{figure}
%\mbox{}\\[-2.5em]

%%%%%%%%%%%%%%%%%%%%%%%%%%%%%%%%%%%%%%%%%%%%
\section{A small but nonzero cosmological constant and the scalar-tensor theory of gravity}
%%%%%%%%%%%%%%%%%%%%%%%%%%%%%%%%%%%%%%%%%%

According to the recent detailed analyses of the Type Ia SN's our universe is ``accelerating," best described by a small but nonzero positive cosmological constant, with the size expressed by $\Omega_\Lambda \equiv \Lambda/\rho_{\rm crt} \sim 0.7$.  The critical density $\rho_{\rm crt} = \frac{3}{8\pi G}H^2$  shows an over-all  behavior as $t^{-2}$ with the coefficient which is of the order one in  the Planckian unit system defined by $c= \hbar = \MP (=(8\pi G)^{-1/2}) =1$.  Note that  the present age of the universe $t_0 \sim 14 {\rm Gy}$ corresponds to $\sim 10^{60}$ in this unit system.  We then find today's value of $\rho_{\rm crt}\sim \Lambda \sim 10^{-120}$.  In contrast, we have a theoretical  estimate $\Lambda_{\rm th}$ naturally  $\sim 1$ in this unit system.  The discrepancy of 120 orders of magnitude will be understood probably only by the ``scenario of a decaying cosmological constant:" $\Lambda$ is not a true constant, but decays like $t^{-2}$.  One of the ways to implement this scenario is provided by the {\it scalar-tensor theory of gravity}.

The theory features the well-known nonminimal coupling term $(1/2)\sqrt{-g}\xi \phi^2 R$ that replaces the standard Einstein-Hilbert term, where $4 \xi = | \omega |^{-1}$ shows how strongly the scalar field $\phi$ couples with matter.  The size of $\omega$ has been best constrained by the solar-system experiments; $\omega \gsim 3600$, or $\xi\lsim 0.7\times 10^{-4}$, though a theoretically natural estimate is
of the order 1.  For example, string theory in its simplest interpretation gives  $\omega = -1$, in contradiction with the observations.  But the solar-system experiments may not constrain the realistic scalar-tensor theory, because the scalar field is likely massive, and the scalar force does not reach the distances of the solar system.

Incidentally,  string theory offers a nice place for the scalar field to live comfortably as a spinless companion of the metric tensor, and is called a ``dilaton."

A conformal transformation $g_{\mu\nu}\rightarrow g_{*\mu\nu}=\Omega^2(x)g_{\mu\nu}$ is crucially important in any version of the scalar-tensor theory, and in any theory of gravity in which a scalar field plays a role.  Also as a terminology, under this transformation we say, we move from one conformal frame (CF) to another.  It then follows that physics looks different from frame to frame.   So the issue is how we can single out a correct and physical CF out of infinitely many of them. At this moment, we  talk about  only two types of them; the J(ordan) frame and the E(instein) frame.

The J frame is the CF in which the Lagrangian takes the form with a  nonminimal coupling term.  String theory is formulated in this type of CF.  For this reason,  the J frame might be called a string CF, or sometimes  a theoretical CF.  On the other hand, in the E frame the nonminimal coupling term has been transformed  to the standard EH term.  It then seems reasonable to start with assuming a truly constant $\Lambda$ term included in the J frame Lagrangian, and to try to see what physics looks like in the E frame, expected to be a physical frame.

Note also that in the E frame the {\em canonical} scalar field is $\sigma$ related to $\phi$ by $\phi\sim e^{\zeta\sigma}$ where $\zeta^{-2}=6+4\omega$, and the purely constant term of $\Lambda$ in the J frame is now converted to a term in the E frame; $\Lambda e^{-4\zeta\sigma}$.

We then set up the cosmological equations in the Friedmann universe with $k=0$.  We work out the solutions, particularly the attractor solution in the E frame.  In this solution, the energy density  $\rho_\sigma$ of the scalar field acts as $\Lambda_{\rm eff}$, the effective cosmological constant in the E frame, and is shown to decay like $t_*^{-2}$, where $t_*$ is the cosmic time in the E frame.  This is the way we can implement the scenario of a decaying $\Lambda$.  However, this smooth and monotonic behavior is not enough to understand the observed acceleration, which requires us to come up with something that behaves like a constant.  To meet this requirement is not an easy task, as long as we stick  to the idea of the decaying $\Lambda$.  After some desperate efforts, we came to propose to introduce another scalar field $\chi$ together with its specific interaction.  Skipping the details altogether at this moment, we show the result [\cite{twosc},\cite{yfkm}].

%%%%%%%%%%%%%%%%%%%%% fig. 3 %%%%%%%%%%%%%%%%%

\begin{figure}[t]
\hspace*{2.cm}
\parbox[b]{6.0cm}
{
\epsfxsize=4.5cm
\epsffile{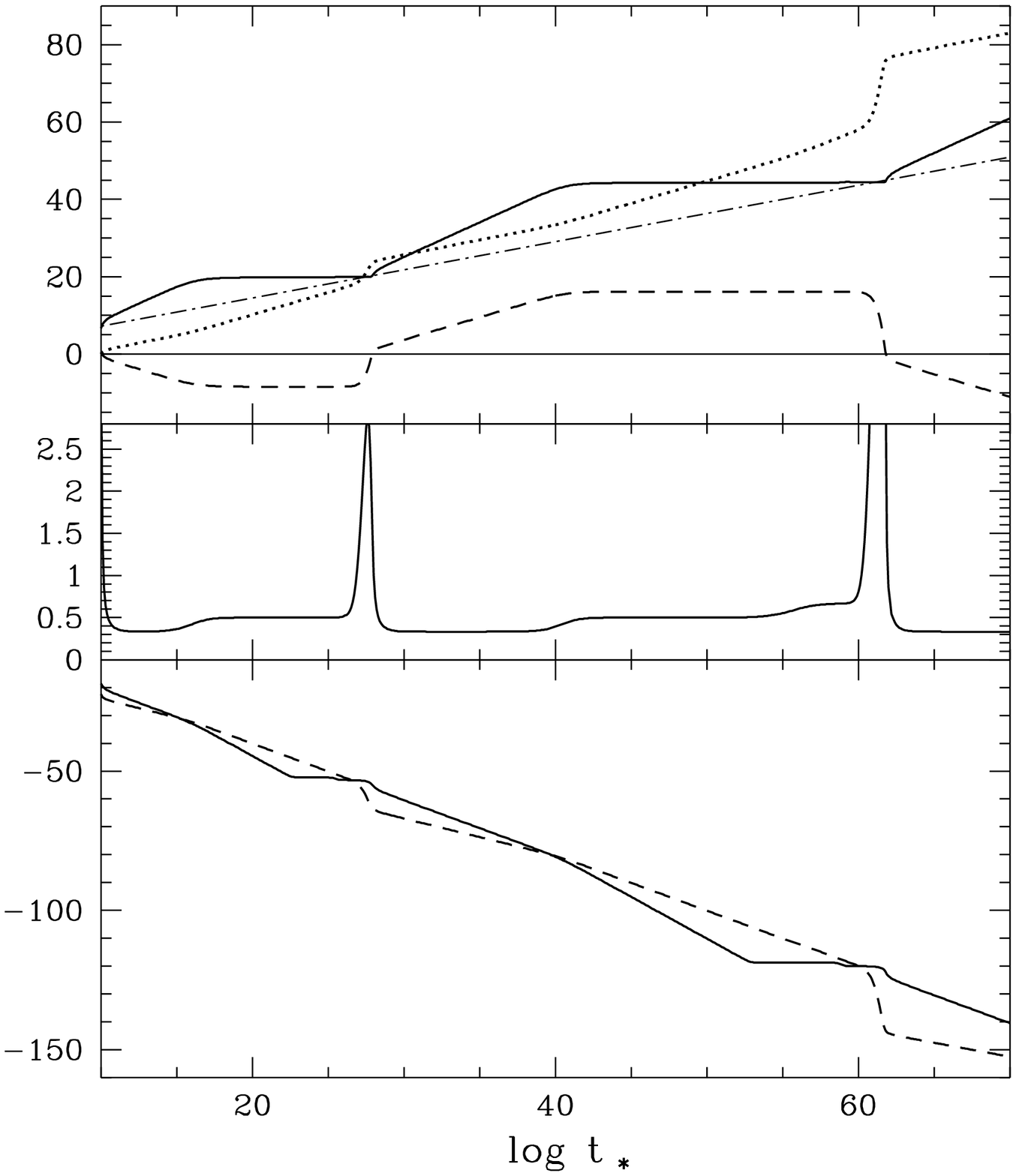}
\mbox{}\\[-1.7em]
\hspace*{6.2em}{\scriptsize Fig. 3a}
}
\parbox[b]{4.9cm}
{
\epsfxsize=4.5cm
\epsffile{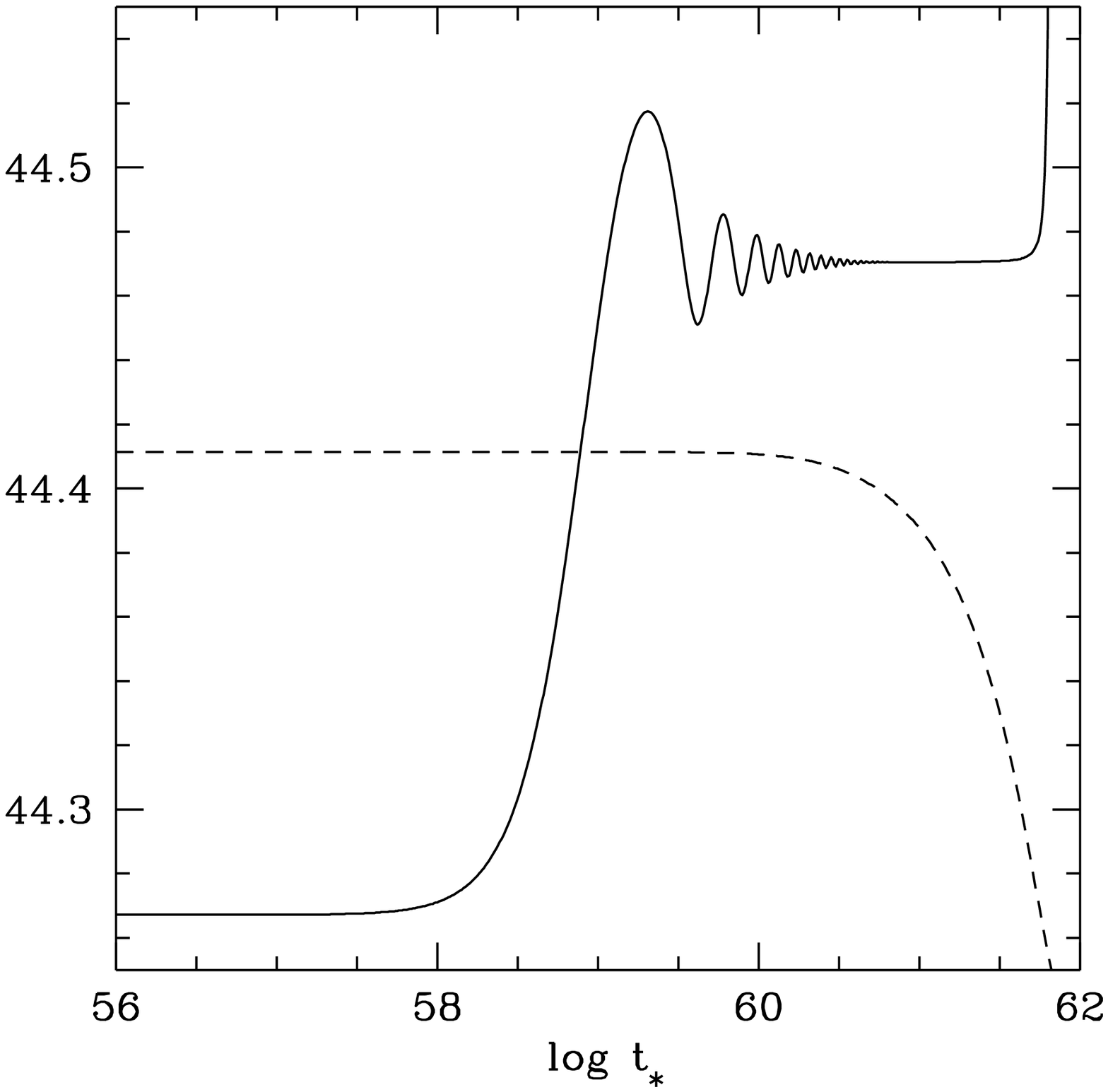}
\mbox{}\\[-2.2em]
\hspace*{6em}{\scriptsize Fig. 3b}
}
\end{figure}

Fig. 3a is an example of our solutions in the E frame.  The horizontal axis is $\log t_*$ in the Planckian unit system; the present epoch is somewhere around 60.  Both of the ordinary matter energy density $\rho_*$ (dashed curve in the lower diagram), including dark matter, and $\Lambda_{\rm eff}= \rho_s$ (solid curve) for  the energy density of the scalar fields, $\sigma$ and $\chi$, fall off $\sim t_*^{-2}$ as an overall behavior, hence respecting the scenario of decaying $\Lambda$.  However, we find occasional deviations from $t_*^{-2}$, particularly  the plateau behavior of $\rho_s$.  A nearly constant $\rho_s$ mimics a cosmological constant, but it is remarkable that the plateau crosses $\rho_*$, corresponding to $\Omega_\Lambda =0.5$.  Nearly in coincidence with this crossing, the scale factor (dotted curve for its logarithm in the upper diagram) shows acceleration, fitting the observed result.  But this ``mini-inflation" does not last forever; it dies out soon on this time scale.

This crossing is a unique feature of the dynamics of the two interacting scalar fields, also indicated by the sudden changes of $\sigma$ and $\chi$ (solid and dashed curves, respectively, in the upper diagram).  We focus particularly on a small portion in the upper diagram, showing a sudden but simple change of $\sigma$ around the present epoch.  We magnify it by the ratio 330 times in the vertical direction.

The result is Fig. 3b showing a damped oscillation of $\sigma$, which, like a hidden secret, is visible only after the huge magnification.  This is a ``heart" of the underlying mechanism designed for a small but nonzero $\Lambda$.  The oscillation itself is so small in the amplitude that its details rarely affect the way of acceleration in practice.  But it may show up somewhere else, particularly through the way $\alpha$ changes with time.

%%%%%%%%%%%%%%%%%%%%%%%%%%%%%%%%%%%%
\section{Varying $\alpha$}
%%%%%%%%%%%%%%%%%%%%%%%%%%%%%%%%%%%%
We have derived the following relation between $\Delta\sigma$ and $\Delta\alpha$:
%%%%%%%%%%%%%%%%%%%%%%
\beq
\frac{\Delta\alpha}{\alpha}={\cal Z}\frac{\alpha}{2\pi}\zeta \Delta\sigma,
\label{pt_21}
\eeq
where ${\cal Z}$ is a coefficient of the order 1.

To be substituted into RHS, we take $\sigma$ tentatively from the above solution, and compute the LHS, plotted in Fig. 2b against the fractional lookback time.  When compared with the observation, the lower diagram of Fig. 2a,   the agreement appears rather poor. We have to try other initial values and parameters for better fits.  It may take some time, though we are sure that we can do this without affecting much the way the universe is accelerated.  In other words, the theory is constrained more in detail by the time-variability of $\alpha$ than by the acceleration of the universe.  We still emphasize that we are coming close to understanding  the two different kinds of phenomena in terms of a single origin, the dynamical scalar field.  The most crucial here was \reflef{pt_21}).  We outline briefly how we reached  this result [\cite{yfkm}].

\begin{itemize}
\item It was necessary to modify the simplest Brans-Dicke model to be consistent with standard cosmology with $\Lambda$ included, a highly non-trivial task.  We also followed the scenario that scale invariance is broken first spontaneously so that $\sigma$ is a Nambu-Goldstone boson, which is massless and is duly called a ``dilaton."  The symmetry is then broken explicitly, with $\sigma$ now as a pseudo NG boson, which is naturally massive.  In this way, we have $\sigma$ finally as non-Newtonian gravity, featuring small violation of WEP.
\item From a technical point of view we started with assuming a purely constant $\alpha$ in the J frame, then finding a $\sigma$-dependence in the E frame.  The  basic tool was QED.  Interesting enough the result \reflef{pt_21}) is finite, rather than divergent, in a way of {\em quantum anomaly,} an intriguing concept in the relativistic quantum field theory.
\item Finally a detailed study on the choice of CF gives $\Delta G/G = -9 {\cal Z}^{-1} (\Delta \alpha/\alpha)$, and hence $ |\dot{G}/G| \lsim 10^{-(14-15)}{\rm y}^{-1}$, somewhat below the current upper bounds.

\end{itemize}
\mbox{}\\
\bcent
{\large\bf References}
\ecent
\begin{enumerate}
{\small 
\item\label{shly}A.I. Shlyakhter: Nature {\bf 264} (1976) 340: ATOMKI Report A/1 (1983), \vspace{-.8em}unpublished.
\item\label{yf8}Y. Fujii, A. Iwamoto, T. Fukahori, T. Ohnuki,
M. Nakagawa, H. Hidaka, Y. Oura and P. M\"{o}ller: Nucl. Phys. {\bf
B573} (2000) \vspace{-.8em}377: See also hep-ph/0205206.
\item\label{dd}T. Damour and F. Dyson: Nucl. Phys. {\bf B480} (1996) \vspace{-.8em}37.
\item\label{webb}J.K. Webb, M.T. Murphy, V.V. Flambaum, V.A. Dzuba, J.D. Barrow, C.W. Churchill, J.X. Prochaska and A.M. Wolfe, Phys. Rev. Lett. {\bf
87} (2001) 091301\vspace{-.8em}.
\item\label{twosc}Y. Fujii, Phys. Rev. {\bf D62} (2000) 064004\vspace{-1.0em}.
\item\label{yfkm}Y. Fujii and K. Maeda, {\it The scalar-tensor theory of gravitation}, Cambridge University \vspace{-.4em}Press (to be published). 
}
\end{enumerate}

%%%%%%%%%%%%%%%%%%%%%%%%%% figs %%%%%%%%%%%%%%%%%%%%%%%%%%%%

\end{document}